\documentclass[showpacs,showkeys,eqsecnum,prd,aps,twoside,twocolumn]{revtex4-1}
\usepackage{amsfonts,amsmath,amssymb,bm}

\begin{document}

\title{Canonical quantization, path integral representations, and
pseudoclassical description of massive Weyl neutrinos in external backgrounds%
}
\author{Maxim Dvornikov${}^{a,b}$}
\email{maxim.dvornikov@usp.br}
\author{D.~M.~Gitman${}^{a}$}
\email{gitman@if.usp.br}
\affiliation{${}^{a}$Institute of Physics, University of S\~{a}o Paulo, CP 66318, CEP
05315-970 S\~{a}o Paulo, SP, Brazil;\\
${}^{b}$Pushkov Institute of Terrestrial Magnetism, Ionosphere and Radiowave
Propagation (IZMIRAN), \\ 142190 Moscow, Troitsk, Russia}
\date{\today}

\begin{abstract}
We study massive $1/2$-spin particles in various external backgrounds
keeping in mind applications to neutrino physics. We are mainly interested
in massive Majorana (Weyl) fields. However, massive neutral Dirac particles
are also considered. We formulate classical Lagrangian theory of the massive
Weyl field in terms of Grassmann-odd two-component spinors. Then we
construct the Hamiltonian formulation of such a theory, which turns out to
be a theory with second-class constraints. Using this formulation we
canonically quantize the massive free Weyl field. We derive propagators of
the Weyl field and relate them to the propagator of a massive Dirac
particle. We also study the massive Weyl particles propagating in the
background mater. We find the path integral representation for the
propagator of such a field, as well as the corresponding pseudoclassical
particle action. The massless limit of the Weyl field interacting with the
matter is considered and compared with results of other works. Finally, the
path integral representation for the propagator of the neutral massive Dirac
particle with an anomalous magnetic moment moving in the background matter
and external electromagnetic field, as well as the corresponding
pseudoclassical particle action are constructed.
\end{abstract}

\pacs{11.10.Ef, 03.65.Pm, 14.60.Pq, 14.60.St}
\keywords{massive Majorana and Dirac neutrinos; Weyl field; pseudoclassical
action; background matter; external electromagnetic field}

\maketitle

\section{Introduction\label{sec:INTR}}

The study of neutrinos is one of the fastest developing areas of
elementary particle physics. This is mainly due to recent experimental
achievements in measuring the parameters of the neutrino mixing matrix~\cite{An12}. Great experimental efforts have been devoted to clarify the question of
whether neutrinos are Dirac or Majorana particles (see, e.g., Ref.~\cite{Aug12}). Nevertheless the problem of the neutrino's nature still remains
open. It should be noted that the Majorana type of neutrino masses are predicted
in some models of the neutrino mass generation, like the see-saw mechanism~\cite{MohSmi06}. However, one cannot exclude the possibility that neutrinos
will turn out to be Dirac particles. In the present work, we study, in the main (except
Sec.~\ref{sec:DIRAC}), Majorana (or Weyl) fields representing neutrinos.

The first quantum field theory treatment of massive Majorana neutrinos was
made in Ref.~\cite{Cas57}, where it was proposed to describe a massive
Majorana field in terms of two-component Weyl spinors. Some processes
involving neutrinos, like $\beta $ decay, were also discussed in Ref.~\cite{Cas57} using Weyl field formalism. Mixed massive Majorana neutrinos
possessing anomalous transition magnetic moments in an external magnetic
field were studied in Ref.~\cite{SchVal81}. The treatment of massive
Majorana fermions based of spinors, which belong to the nonstandard Wigner
classes and violate the Lorentz invariance, was proposed in Ref.~\cite{AhlLeeSch11}. Recently a detailed quantization of massive Majorana
particles, described in terms of $c$-number Weyl fields, in vacuum and in
background matter, was considered in Ref.~\cite{Dvo12}.

In the present work, we continue the rigorous study of the massive Weyl field
in vacuum as well as in background matter in the framework of classical and
quantum theory. In Sec.~\ref{sec:CLASS}, we start with the construction of
the classical Lagrangian and Hamiltonian formulations of the massive Weyl
field in a vacuum using the Grassmann-valued two-component spinors. Then,
based on the constructed Hamiltonian formulation, we canonically quantize
this theory. We also introduce propagators of the massive Weyl field and
relate them with the corresponding propagators of the massive Dirac field.
The interaction of the system of spin-$1/2$ fermionic fields with general
external fields is described in Sec.~\ref{sec:FERMEXTF}. Then, in Sec.~\ref{sec:MAJINTMAT}, we discuss the propagation of a massive Weyl field in a
background matter. We find the path integral representations for all the
propagators and derive the corresponding pseudoclassical actions; in
particular, the pseudoclassical action for massive Weyl particles moving in a
background matter. The massless limit of a Weyl field interacting with
background matter is also considered. Finally, in Sec.~\ref{sec:DIRAC}, we
briefly discuss the massive neutral Dirac particles possessing an anomalous
magnetic moment, moving in background matter and in an external
electromagnetic field. We summarize the obtained results in Sec.~\ref{sec:CONCL}.

\section{Classical and quantum theory of a free massive Weyl field \label{sec:CLASS}}

As we mentioned in Sec.~\ref{sec:INTR}, the most prominent candidates to be
described in terms of Majorana fields are neutrinos. Now, it is a well-established fact that there are three active neutrino generations (see,
e.g., Ref.~\cite{Abd05}) and one has a nonzero mixing between them. In some
theoretical models a number of sterile neutrinos is also predicted~\cite{Vol02}. After the diagonalization of the neutrino mass matrix, one gets a
number of fermionic fields which are Majorana~\cite{Kob80}. Therefore, if we
deal with Majorana particles in vacuum, we can study a single massive field
without loss of generality. Our results can be straightforwardly generalized
to include several neutrino generations.

A fermionic spin-$1/2$ field $\psi \left( x\right) $ of the mass $m$ obeys
the Dirac equation:
\begin{equation}\label{eq:Direq}
  (\mathrm{i}\gamma^{\mu}\partial_{\mu}-m)\psi
  \left(
    x
  \right) =0,
  \quad
  x=
  \left(
    x^{\mu}
  \right) =
  \left(
    x^{0},\mathbf{x}
  \right).
\end{equation}
For our purposes, it convenient to chose the Dirac $\gamma $ matrices in the
chiral representation:
\begin{gather}  \label{Diracmatr}
  \gamma^{\mu}=
  \left(
    \begin{array}{cc}
      0 & -\sigma^{\mu} \\
      -\tilde{\sigma}^{\mu} & 0%
    \end{array}%
  \right),
  \notag
  \\
  \sigma^{\mu}=(\sigma^{0},-\boldsymbol{\sigma}),
  \quad
  \tilde{\sigma}^{\mu}=(\sigma^{0},\boldsymbol{\sigma}),
\end{gather}
where $\sigma^{0}=\sigma_{0}$ is the unit $2\times 2$ matrix, and $\boldsymbol{\sigma}$ are the Pauli matrices (see, e.g., Ref.~\cite{ItzZub80}).

Let us represent the bispinor $\psi$ in terms of two-component Weyl spinor $\eta$ as follows:
\begin{equation}\label{eq:Majspinor}
  \psi =
  \left(
    \begin{array}{c}
      \mathrm{i}\sigma_{2}\eta^{\ast} \\
      \eta%
    \end{array}%
  \right).
\end{equation}
Note that such a bispinor satisfies the Majorana condition $\psi^{c}=\mathrm{i}\gamma^{2}\psi^{\ast}=\psi $. The Dirac equation~(\ref{eq:Direq}) implies the following equation for the spinor $\eta $,
\begin{equation}\label{eq:Weyleq}
  \sigma^{\mu}\partial_{\mu}\eta +m\sigma_{2}\eta^{\ast}=0;
\end{equation}
see, e.g. Ref.~\cite{FukYan03}.

Note that Eq.~(\ref{eq:Weyleq}) is the Euler-Lagrange equation for the
Lagrangian
\begin{equation}\label{eq:WeylLagr}
  \mathcal{L}=
  \mathrm{i}\eta^{\dagger}\sigma^{\mu}\partial_{\mu}\eta -
  \frac{\mathrm{i}}{2}m
  \left(
    \eta^{\mathrm{T}}\sigma_{2}\eta -\eta^{\dagger}\sigma_{2}\eta^{\ast}
  \right).
\end{equation}
In this Lagrangian, we treat $\eta $ as Grassmann-odd fields~\cite{Berez66}.
In particular, due to this reason the mass term on the right-hand side of
Eq.~(\ref{eq:WeylLagr}) is not zero.

In case we study massless spin-$1/2$ particles, the corresponding bispinor $\psi_{0}$ satisfies the equation
\begin{equation}\label{eq:Direqm0}
  \mathrm{i}\gamma^{\mu}\partial_{\mu}\psi_{0}=0.
\end{equation}
It should be noted that there are solutions of Eq.~(\ref{eq:Direqm0}) which
satisfy the conditions $P_{\mathrm{R}}\psi_{0}=0$ (left-handed bispinors)
or $P_{\mathrm{L}}\psi_{0}=0$ (right-handed bispinors), where
\begin{equation}\label{Proj}
  P_{\mathrm{R,L}} = (1\pm \gamma^{5})/2,
  \quad
  \gamma^{5}=\mathrm{i}\gamma^{0}\gamma^{2}\gamma^{2}\gamma^{3}.
\end{equation}
Considering, e.g., a left-handed bispinor, we can present it as
\begin{equation}\label{eq:Majspinorm0}
  \psi_{0}=
  \left(
    \begin{array}{c}
      0 \\
      \eta_{0}%
    \end{array}%
  \right).
\end{equation}
Combining Eqs.~(\ref{eq:Direqm0}) and~(\ref{eq:Majspinorm0}), we get the wave
equation for the spinor $\eta_{0}$,
\begin{equation}  \label{Lagreta0}
  \sigma^{\mu}\partial_{\mu}\eta_{0}=0.
\end{equation}
The latter equation is the Euler-Lagrange equation for the following
Lagrangian:
\begin{equation}\label{eq:WeylLagrm0}
  \mathcal{L}_{0}=\mathrm{i}\eta_{0}^{\dagger}\sigma^{\mu}\partial_{\mu}\eta_{0}.
\end{equation}

Starting with the Lagrangian in Eq.~(\ref{eq:WeylLagr}), we find the canonical
momenta as
\begin{equation}\label{eq:canmom}
  \pi =\frac{\partial_{r}\mathcal{L}}{\partial \dot{\eta}} =
  \mathrm{i}\eta^{\ast},
  \quad
  \pi^{\ast}=\frac{\partial_{r}\mathcal{L}}{\partial \dot{\eta}^{\ast}}=0,
\end{equation}
where the subscript $r$ denotes the right derivatives; see Ref.~\cite{GitTyt90}. One can easily see that the system in question has two
second-class constraints:
\begin{equation}\label{eq:constr}
  \Phi_{1}=\pi -\mathrm{i}\eta^{\ast}=0,
  \quad
  \Phi_{2}=\pi^{\ast}=0.
\end{equation}
The extended Hamiltonian reads
\begin{equation}\label{eq:extHam}
  \mathcal{H}_{1}=\mathcal{H}+\Phi_{1}\lambda_{1}+\lambda_{2}\Phi_{2},
\end{equation}
where
\begin{equation}\label{eq:Ham}
  \mathcal{H}=\mathrm{i}\eta^{\dagger}(\boldsymbol{\sigma \nabla})\eta +
  \frac{\mathrm{i}}{2}m\left( \eta^{\mathrm{T}}\sigma_{2}\eta -
  \eta^{\dagger}\sigma_{2}\eta^{\ast}\right),
\end{equation}
and $\lambda_{1,2}$ are Lagrange multipliers which are Grassmann-odd
fields. These multipliers can be found explicitly,
\begin{equation*}
  \lambda_{1}=(\boldsymbol{\sigma \nabla})\eta -m\sigma_{2}\eta^{\ast},
  \quad
  \lambda_{2}=
  -
  \left(
    \boldsymbol{\nabla}\eta^{\dagger}
  \right)
  \boldsymbol{\sigma}+m\eta^{\mathrm{T}}\sigma_{2},
\end{equation*}
from the conditions of the constraint conservation
\begin{equation}\label{eq:constcons}
  \left\{
    \Phi_{1},\mathcal{H}_{1}
  \right\} =
  \left\{
    \Phi_{2},\mathcal{H}_{1}
  \right\} =0,
\end{equation}
where $\left\{\ \right\} $ are the generalized Poisson brackets, see
Ref.~\cite{GitTyt90}.

The presence of the second-class constraints in a system implies that in the
equations of motion and in the quantization procedure, the generalized
Poisson brackets must be replaced by the generalized Dirac brackets, defined
as
\begin{equation}\label{eq:Dirbrac}
  \left\{
    F,G
  \right\}_{\mathrm{D}} =
  \left\{
    F,G
  \right\} -
  \left\{
    F,\Phi_{i}
  \right\}
  C_{ij}
  \left\{
    \Phi_{j},G
  \right\},
\end{equation}
where the matrix $(C_{ij})$ has the components $C_{ij}=\left\{ \Phi_{i},\Phi_{j}\right\}^{-1}$. In our case,
\begin{equation}\label{eq:Cdef}
  C =
  \begin{pmatrix}
    0 & \mathrm{i} \\
    \mathrm{i} & 0
  \end{pmatrix}
  \delta^{3}(\mathbf{x}-\mathbf{y}).
\end{equation}
The nonzero Dirac brackets of the basis fields have the form
\begin{align}\label{eq:fundDirbrac}
  &
  \left\{
    \eta (\mathbf{x},t),\eta^{\ast}(\mathbf{y},t)
  \right\}_{\mathrm{D}} =
  -\mathrm{i}\delta^{3}(\mathbf{x}-\mathbf{y}),
  \notag
  \\
  &
  \left\{
    \eta (\mathbf{x},t),\pi (\mathbf{y},t)
  \right\}_{\mathrm{D}} =
  \delta^{3}(\mathbf{x}-\mathbf{y}).
\end{align}
Hamiltonian equations of motion have the form
\begin{align}\label{eq:Weyleqfull}
  &
  \dot{\eta}=
  \left\{
    \eta ,\mathcal{H}
  \right\}_{\mathrm{D}} =
  (\boldsymbol{\sigma \nabla})\eta -m\sigma_{2}\eta^{\ast},
  \notag
  \\
  &
  \dot{\eta}^{\ast} =
  \left\{
    \eta^{\ast},\mathcal{H}
  \right\}_{\mathrm{D}} =
  \left(
    \boldsymbol{\sigma}^{\mathrm{T}}\boldsymbol{\nabla}
  \right)
  \eta^{\ast}+m\sigma_{2}\eta,
  \notag
  \\
  &
  \dot{\pi}=
  \left\{
    \pi ,\mathcal{H}
  \right\}_{\mathrm{D}} =
  \mathrm{i}
  \left(
    \boldsymbol{\sigma}^{\mathrm{T}}\boldsymbol{\nabla}
  \right)
  \eta^{\ast} + \mathrm{i}m\sigma_{2}\eta,
  \notag
  \\
  &
  \dot{\pi}^{\ast} =
  \left\{
    \pi^{\ast},\mathcal{H}
  \right\}_{\mathrm{D}}=0.
\end{align}
Of course, one can see from Eq.~(\ref{eq:Weyleqfull}) that the evolution
equation for $\eta $ coincides with Eq.~(\ref{eq:Weyleq}).

Using Eq.~(\ref{eq:constr}), it is convenient to rewrite the wave equations
for $\eta $ and $\pi $ as follows:
\begin{align}\label{eq:Weyleqetapi}
  \dot{\eta} & =
  (\boldsymbol{\sigma \nabla})\eta +\mathrm{i}\sigma_{2}m\pi,
  \notag
  \\
  \dot{\pi} & =
  \left(
    \boldsymbol{\sigma}^{\mathrm{T}}\boldsymbol{\nabla}
  \right)
  \pi +\mathrm{i}\sigma_{2}m\eta,
\end{align}
where the operation of the complex conjugation is excluded.

The two-component Weyl spinor field $\eta $ is fermionic. When doing the
canonical quantization of such a field, one should replace $\eta $ and $\pi $
with the operators $\eta \rightarrow \hat{\eta}$ and $\pi \rightarrow \hat{\pi}$,
and define the equal-time anticommutators for these operators via the
corresponding Dirac brackets (see Ref.~\cite{GitTyt90}). The nonzero equal-time anticommutators for the basic Heisenberg operators are
\begin{align}\label{eq:canquantfundanticom}
  \left[
    \hat{\eta}(\mathbf{x},t),\hat{\pi}(\mathbf{y},t)
  \right]_{+} = & \mathrm{i}
  \left.
    \left\{
      \eta (\mathbf{x},t),\pi (\mathbf{y},t)
    \right\}_{\mathrm{D}}
  \right\vert_{\eta = \hat{\eta},\pi =\hat{\pi}}
  \notag
  \\
  & =
  \mathrm{i}\delta^{3}(\mathbf{x}-\mathbf{y}),
\end{align}
where the fundamental Dirac bracket is given in Eq.~(\ref{eq:fundDirbrac}).
Equation~(\ref{eq:constr}) holds true for the corresponding operators, such that
we can rewrite Eq.~(\ref{eq:canquantfundanticom}) as follows:
\begin{equation}\label{eq:anticommod}
  \left[
    \hat{\eta}(t,\mathbf{x}),\hat{\eta}^{\dagger}(t,\mathbf{y})
  \right]_{+} =
  \delta^{3}(\mathbf{x}-\mathbf{y}).
\end{equation}

Constructing the Heisenberg operators $\hat{\eta}(x)$ that satisfy Eq.~(\ref{eq:Weyleq}) and the anticommutation relation [Eq.~(\ref{eq:anticommod})], we
obtain
\begin{align}\label{eq:solWeyleq}
  \hat{\eta}(x)= &
  \int \frac{\mathrm{d}^{3}\mathbf{p}}{(2\pi)^{3/2}}
  \sqrt{\frac{E+|\mathbf{p}|}{2E}}
  \notag
  \\
  & \times
  \bigg[
    \left(
      \hat{a}_{-}w_{-} -
      \frac{m}{E+|\mathbf{p}|}\hat{a}_{+}w_{+}
    \right)
    e^{-\mathrm{i}px}
    \notag
    \\
    & +
    \left(
      \hat{a}_{+}^{\dagger}w_{-} +
      \frac{m}{E+|\mathbf{p}|}\hat{a}_{-}^{\dagger}w_{+}
    \right)
    e^{\mathrm{i}px}
  \bigg],
\end{align}
where $E=\sqrt{|\mathbf{p}|^{2}+m^{2}}$;
\begin{equation*}
  w_{+} =
  \left(
    \begin{array}{c}
      e^{-\mathrm{i}\phi /2}\cos \theta /2 \\
      e^{\mathrm{i}\phi /2}\sin \theta /2%
    \end{array}%
  \right),
  \quad
  w_{-}=
  \left(
    \begin{array}{c}
      -e^{-\mathrm{i}\phi /2}\sin \theta /2 \\
      e^{\mathrm{i}\phi /2}\cos \theta /2%
    \end{array}%
  \right) 
\end{equation*}
are chiral amplitudes; the angles $\phi $ and $\theta $ fix the direction of
the particle momentum, $\mathbf{p}=|\mathbf{p}|(\cos \phi \sin \theta ,\sin
\phi \sin \theta ,\cos \theta )$; and $\hat{a}_{\pm}^{\dagger}(\mathbf{p})$
and $\hat{a}_{\pm}(\mathbf{p})$ are the creation and annihilation
operators,
\begin{align}
  \left[
    \hat{a}_{\sigma}(\mathbf{k}),
    \hat{a}_{\sigma^{\prime}}^{\dagger}(\mathbf{k}^{\prime})
  \right]_{+} = &
  \delta_{\sigma \sigma^{\prime}}\delta^{3}(\mathbf{k}-\mathbf{k}^{\prime}),
  \notag
  \\
  \left[
    \hat{a}_{\sigma}(\mathbf{k}),
    \hat{a}_{\sigma^{\prime}}(\mathbf{k}^{\prime})
  \right]_{+} = & 0,
  \notag
  \\
  \left[
    \hat{a}_{\sigma}^{\dagger}(\mathbf{k}),
    \hat{a}_{\sigma^{\prime}}^{\dagger}(\mathbf{k}^{\prime})
  \right]_{+} = & 0,
  \quad
  \sigma =\pm{}.
\label{eq:cananticomaad}
\end{align}
Note that Eqs.~(\ref{eq:solWeyleq}) and~(\ref{eq:cananticomaad}) coincide
with analogous expressions derived in Ref.~\cite{FukYan03} using heuristic
arguments.

Using Eqs.~(\ref{eq:Ham}), (\ref{eq:solWeyleq}), and~(\ref{eq:cananticomaad}), we get the total energy of the massive Weyl field via
the creation and annihilation operators as follows:
\begin{align}\label{energy}
  E_{\mathrm{tot}} = &
  \int \mathrm{d}^{3}\mathbf{r}\mathcal{H} =
  \int \mathrm{d}^{3}\mathbf{p}E
  \left(
    \hat{a}_{-}^{\dagger}\hat{a}_{-}+\hat{a}_{+}^{\dagger}\hat{a}_{+}
  \right)
  \notag
  \\
  & +
  \text{\textrm{divergent terms}},
\end{align}
where ``divergent terms" contains $\delta^{3}(0)$ and can be removed by the normal ordering of operators. This result
confirms the chosen interpretation of the operators $\hat{a}_{\pm}^{\dagger}(\mathbf{p})$ and $\hat{a}_{\pm}(\mathbf{p}).$

Let us introduce two propagators of the massive Weyl field in vacuum as
follows
\begin{align}\label{WeylProp}
  S_{1}(x-y) = &
  \mathrm{i}
  \langle
    0|T\{\eta (x)\eta^{\dagger}(y)\}|0
  \rangle,
  \notag
  \\
  S_{2}(x-y) = &
  -
  \langle
    0|T\{\eta (x)\eta^{\mathrm{T}}(y)\}|0
  \rangle.
\end{align}
Here $|0\rangle $ is the vacuum vector for the annihilation operators $\hat{a}_{\pm}(\mathbf{p})$, and $T$ is the sign of the chronological ordering. With
the help of Eqs.~(\ref{eq:solWeyleq}) and~(\ref{eq:cananticomaad}), we can cast the
functions $S_{1,2}$ into the following forms~\cite{FukYan03}:
\begin{align}\label{S-1,2}
  S_{1}(x) = &
  \int \frac{\mathrm{d}^{4}p}{(2\pi )^{4}}
  \frac{\tilde{\sigma}^{\mu}p_{\mu}}{m^{2}-p^{2}-\mathrm{i}\epsilon}
  e^{-\mathrm{i}px},
  \notag
  \\
  S_{2}(x) = &
  \int \frac{\mathrm{d}^{4}p}{(2\pi )^{4}}
  \frac{\sigma_{2}m}{m^{2}-p^{2}-\mathrm{i}\epsilon}
  e^{-\mathrm{i}px},
\end{align}
where $\epsilon $ is a positive, infinitesimally small constant, and the
matrices $\tilde{\sigma}^{\mu}$ are defined in Eq.~(\ref{Diracmatr}).

For our purposes, it is convenient to relate the introduced propagators [Eq.~(\ref{S-1,2})] to the propagator of the massive Dirac field. To proceed with the
problem, let us introduce the Hamiltonian Weyl field $\Xi $ that is the pair
$\pi $ and $\eta $,
\begin{equation}\label{WFH}
  \Xi =
  \left(
    \begin{array}{c}
      \pi \\
      \eta%
    \end{array}%
  \right).
\end{equation}
Using Eq.~(\ref{eq:Weyleqetapi}), one can find the the Hamiltonian Weyl
field $\Xi $ satisfies the equation $\hat{K}\Xi =0$, where the $4\times 4$
matrix operator $\hat{K}$ reads
\begin{equation}\label{eq:Kdef}
  \hat{K} =
  \left(
    \begin{array}{cc}
      \partial_{t}-(\boldsymbol{\sigma}^{\mathrm{T}}\boldsymbol{\nabla}) &
      -\mathrm{i}\sigma_{2}m \\
      -\mathrm{i}\sigma_{2}m & \partial_{t}-(\boldsymbol{\sigma \nabla})
    \end{array}
  \right).
\end{equation}
The causal propagator for the Hamiltonian field $\Xi$ is
\begin{align}\label{CausalProp}
  S_{\mathrm{c}}(x-y) = &
  \langle
    0|T\{\Xi (x)\Xi^{\dagger}(y)\}|0
  \rangle,
  \notag
  \\
  \hat{K}S_\mathrm{c}(x) = & -\delta^{4}(x).
\end{align}
Introducing the function $\tilde{S}_\mathrm{c}$ instead of $S_{\mathrm{c}}$,
\begin{equation*}
  \tilde{S}_{\mathrm{c}} = -\mathfrak{S}_{2} S_\mathrm{c} \mathfrak{S}_{2}
  \gamma^{0}\gamma^{5},
  \quad
  \mathfrak{S}_{2} = \text{diag}(\sigma_{2},\sigma_{0}),
\end{equation*}
we derive for this function the following equation:
\begin{equation}\label{eq:tildeSvac}
  \left(
    \mathrm{i}\Gamma^{\mu}\partial_{\mu}-m\Gamma^{5}
  \right)
  \tilde{S}_{\mathrm{c}} = \delta^{4}(x),
\end{equation}
where $\Gamma^{\mu}=\Gamma^{5}\gamma^{\mu}$ and $\Gamma^{5}=-\mathrm{i}\gamma^{5}$. The $\gamma$ matrices
$\Gamma^{n}=(\Gamma^{\mu},\Gamma^{5}) $ satisfy the usual $\gamma$-matrix algebra
\begin{align}\label{algebra}
  [\Gamma^{k},\Gamma^{n}] = & 2\eta^{kn},
  \quad
  \eta_{kn}=\text{\textrm{diag}}(1,-1,\dots ,-1),
  \notag
  \\
  k,n = & 0,\dots ,3,5,
\end{align}
so that $\tilde{S}_\mathrm{c}$ can be interpreted as the causal propagator
of the massive Dirac field.

The propagators of the massive Weyl field [Eq.~(\ref{S-1,2})] can be related to
the causal propagator $S_\mathrm{c}$ of the Hamiltonian field $\Xi$, and
then to the causal propagator $\tilde{S}_{\mathrm{c}}$ of the massive Dirac
field. First, we write
\begin{align}
  P_{\mathrm{L}}S_{\mathrm{c}}P_{\mathrm{L}} = &
  -\frac{\mathrm{i}}{2}(\sigma_{0}-\sigma_{3})\otimes S_{1},
  \notag
  \\
  P_{\mathrm{L}}S_\mathrm{c}P_{\mathrm{R}} = &
  \frac{\mathrm{i}}{2}(\sigma_{1}-\mathrm{i}\sigma_{2})\otimes S_{2},
  \label{eq:S12}
  \\
  \frac{1}{2}(\sigma_{0}-\sigma_{3})\otimes S_{1} = &
  P_{\mathrm{L}}\tilde{S}_{\mathrm{c}}P_{\mathrm{R}}\Gamma^{0},
  \notag
  \\
  \frac{1}{2}(\sigma_{1}-\mathrm{i}\sigma_{2})\otimes S_{2} = &
  P_{\mathrm{L}}\tilde{S}_{\mathrm{c}}P_{\mathrm{L}}\Gamma^{0}\mathfrak{S}_{2},
  \label{S12tildeS}
\end{align}
where the chiral projectors $P_{\mathrm{L,R}}$ were defined in Eq.~(\ref{Proj}). Then, one can derive from Eqs.~(\ref{eq:S12}) and~(\ref{S12tildeS})
that
\begin{align}\label{eq:S12SVT}
  S_{1} = & -\mathrm{i}\tilde{\sigma}^{\mu}
  \left(
    V_{\mu}-A_{\mu}
  \right),
  \notag
  \\
  S_{2} = & -\mathrm{i}
  \left[
    S-P+\frac{\mathrm{i}}{2}
    \left(
      \tilde{\sigma}^{\mu}\sigma^{\nu}-\tilde{\sigma}^{\nu}\sigma^{\mu}
    \right)
    T_{\mu \nu}
  \right]
  \sigma_{2},
\end{align}
where $S$, $P$, $V_{\mu}$, $A_{\mu}$, and $T_{\mu \nu}$ are the scalar,
pseudoscalar, vector, axial-vector and tensor coefficients in the expansion
of $\tilde{S}_{\mathrm{c}}$ in the independent $\gamma $-matrix basis,
\begin{equation}\label{Scexp}
  \tilde{S}_{\mathrm{c}} =
  S\mathrm{I}+P\gamma^{5}+V_{\mu}\gamma^{\mu}+A_{\mu}\gamma^{5}\gamma^{\mu}+T_{\mu \nu}\sigma^{\mu \nu},
\end{equation}
where $\mathrm{I}$ is the unit $4\times 4$ matrix and $\sigma^{\mu \nu}=(\mathrm{i}/2)[\gamma^{\mu},\gamma^{\nu}]_{-}$. Finally, using Eq.~(\ref{eq:S12SVT}) we can express $S_{1,2}$ as follows:
\begin{eqnarray}
  S_{1} & = &-\frac{\mathrm{i}}{4}\text{tr}
  \left[
    \gamma_{\mu}(1-\gamma^{5})
    \tilde{S}_{\mathrm{c}}
  \right]
  \tilde{\sigma}^{\mu},
  \label{eq:S1tildeS}
  \\
  S_{2} & = &-\frac{\mathrm{i}}{4}\text{tr}
  \left[
    \gamma_{\mu}\gamma^{0}(1-\gamma^{5})\tilde{S}_{\mathrm{c}}
  \right]
  \sigma^{\mu}\sigma_{2}.
  \label{S2tildeS}
\end{eqnarray}
In Sec.~\ref{sec:MAJINTMAT}, we shall give the expression for $\tilde{S}_{%
\mathrm{c}}$ in terms of the path integral.

\section{Interaction of fermions with external fields\label{sec:FERMEXTF}}

In this section we briefly review the interaction of fermionic fields with
external fields. We discuss the general situation when an arbitrary number of
spinor fields are present. The cases of Dirac and Majorana fermions are
compared. We also consider the gravitational interaction of spinor particles.

The most general classical Lagrangian describing the interaction of $1/2$%
-spin fermions $\psi_a$, $a = 1,2,\dots$, which are Grassmann-valued
spinors, with a set of external fields has the form~\cite{DvoStu02},
\begin{align}  \label{Lintgen}
  -\mathcal{L}_\mathrm{int} = &
  s_{ab} \bar{\psi}_a \psi_b + \pi_{ab} \bar{\psi}_a \gamma^5 \psi_b
  \notag
  \\
  & +
  V^\mu_{ab} \bar{\psi}_a \gamma_\mu \psi_b +
  A^\mu_{ab} \bar{\psi}_a \gamma_\mu \gamma^5 \psi_b
  \notag
  \\
  & +
  \frac{1}{2}T^{\mu\nu}_{ab} \bar{\psi}_a \sigma_{\mu\nu} \psi_b +
  \frac{1}{2}\Pi^{\mu\nu}_{ab} \bar{\psi}_a \sigma_{\mu\nu} \gamma^5 \psi_b,
\end{align}
where $s_{ab}$, $\pi_{ab}$, $V^\mu_{ab}$, $A^\mu_{ab}$, $T^{\mu\nu}_{ab}$,
and $\Pi^{\mu\nu}_{ab}$ are the scalar, pseudoscalar, vector, axial-vector,
tensor, and pseudotensor fields, respectively. Note that these external
fields can depend on spatial coordinates.

If we deal with Dirac fermions, the spinors $\psi_a$ and $\bar{\psi}_a$ are
independent degrees of freedom. In this case the external fields in Eq.~(\ref{Lintgen}) are Hermitian matrices in the indices $a$ and $b$.

If we have Majorana fermions, the spinors $\psi_a$ and $\bar{\psi}_a$ are no
longer independent. Instead they obey the relation, cf. Sec.~\ref{sec:CLASS},
\begin{equation}\label{Majcondpsi}
  \psi_a = \mathcal{C} \bar{\psi}^\mathrm{T}_a,
  \quad
  \text{or},
  \quad
  \bar{\psi}_a =
  \psi_a^\mathrm{T} \mathcal{C},
\end{equation}
where $\mathcal{C}$ is the charge conjugation matrix which has the following
properties~\cite{ItzZub80}:
\begin{align}  \label{Cprop}
  & \mathcal{C} = - \mathcal{C}^{-1} = - \mathcal{C}^\mathrm{T} = - \mathcal{C}^\dagger,
  \notag
  \\
  & \mathcal{C} \gamma^\mu \mathcal{C}^{-1} = - \gamma^{\mu\mathrm{T}},
  \quad
  \mathcal{C} \gamma^5 \mathcal{C}^{-1} = \gamma^{5\mathrm{T}},
  \notag
  \\
  & \mathcal{C} \gamma^\mu \gamma^5 \mathcal{C}^{-1} = (\gamma^\mu \gamma^5)^\mathrm{T},
  \quad
  \mathcal{C} \sigma_{\mu\nu} \mathcal{C}^{-1} = - \sigma_{\mu\nu}^\mathrm{T},
\end{align}
which are independent on $\gamma$-matrix representations.

Using Eqs.~(\ref{Lintgen})-(\ref{Cprop}), we immediately find that for
Majorana fermions the matrices $s_{ab}$, $\pi_{ab}$, $A^\mu_{ab}$ should be
symmetric, whereas $V^\mu_{ab}$, $T^{\mu\nu}_{ab}$, and $\Pi^{\mu\nu}_{ab}$
are antisymmetric. This fact, in particular, means that the diagonal (i.e.,
when $a=b$) electromagnetic interaction of Majorana neutrinos vanishes (see
also Sec.~\ref{sec:MAJINTMAT} below).

The interaction of a spinor particle with a gravitational field can be
implemented in a locally Minkowski space by the introduction of the
covariant derivative there (see, e.g., Ref.~\cite{Ham02}). This covariant
derivative also includes the contribution of the torsion, which, in
principle, can have a nonzero value. The Dirac equation in flat
space-time in the presence of torsion was also studied in Ref.~\cite{GeyGitSha00}. It was shown there that the contribution of the torsion is equivalent to
that ofan external axial-vector field [see Eq.~(\ref{Lintgen})]. The issue of whether
there is a difference between the gravitational interaction of Dirac and Majorana
fermions was discussed in Ref.~\cite{SinMobPap06}.

\section{Massive Weyl neutrinos in external fields\label{sec:MAJINTMAT}}

Here we generalize the results of the previous section to include the
interaction of the massive Weyl particles with external backgrounds. We
discuss a particular case of massive neutrinos interacting with background
matter and an external electromagnetic field.

Note that in Sec.~\ref{sec:CLASS} we discussed the case of a single massive
Weyl field. It was also mentioned that a generalization to several mass
eigenstates $\eta_{a}$ with masses $m_{a}$ is not so difficult (see, e.g.,
Sec.~\ref{sec:FERMEXTF}). The most general Lagrangian which involves the
interaction of the fields $\eta_{a}$ with background matter and the
electromagnetic field $(\mathbf{E},\mathbf{B}) $ has the form (see, e.g.,
Refs.~\cite{Kay82,Man88}),
\begin{align}\label{eq:Lagrextfields}
  \mathcal{L} = &
  \mathrm{i}\eta_{a}^{\dagger}\sigma^{\mu}\partial_{\mu}\eta_{a}-\frac{\mathrm{i}}{2}m_{a}
  \left(
    \eta_{a}^{\mathrm{T}}\sigma_{2}\eta_{a}-\eta_{a}^{\dagger}\sigma_{2}\eta_{a}^{\ast}
  \right)
  \notag
  \\
  & -
  g_{ab}^{\mu}\eta_{a}^{\dagger}\sigma_{\mu}\eta_{b}
  \notag
  \\
  &
  -\frac{1}{2}
  \big[
    \mu_{ab}\eta_{a}^{\dagger}\boldsymbol{\sigma}(\mathbf{B}-\mathrm{i}\mathbf{E})\sigma_{2}\eta_{b}^{\ast}
    \notag
    \\
    & +
    (\mu^{\dagger})_{ab}\eta_{a}^{\mathrm{T}}\sigma_{2}(\mathbf{B}+\mathrm{i}\mathbf{E})\boldsymbol{\sigma}\eta_{b}
  \big]
  \notag
  \\
  &
  -\frac{1}{2}
  \big[
    \varepsilon_{ab}\eta_{a}^{\dagger}\boldsymbol{\sigma}(\mathbf{E}+\mathrm{i}\mathbf{B})\sigma_{2}\eta_{b}^{\ast}
    \notag
    \\
    & +
    (\varepsilon^{\dagger})_{ab}\eta_{a}^{\mathrm{T}}\sigma_{2}(\mathbf{E}-\mathrm{i}\mathbf{B})\boldsymbol{\sigma}\eta_{b}
  \big],
\end{align}
where $(\mu_{ab})$ and $(\varepsilon_{ab})$ are the matrices of magnetic
and electric dipole moments.

The interaction with background matter is characterized by the quantities $g_{ab}^{\mu}$, which are Lorentz four-vectors and Hermitian matrices in the
particle species space, $g_{ab}^{\mu \ast}=g_{ba}^{\mu}$. As was shown
in Ref.~\cite{PasSegSemVal00} (see also Sec.~\ref{sec:FERMEXTF}), the
Hermitian matrices $(\mu_{ab})$ and $(\varepsilon_{ab})$ must be
antisymmetric.

The physical implementation of the Lagrangian in Eq.~(\ref{eq:Lagrextfields}) is
a system of massive Majorana neutrinos propagating in background
matter under the influence of an external electromagnetic field. In case we
study the system of massive neutrinos, the components of the matrix $(g_{ab}^{\mu})$ are related to the effective potentials of the active
flavor neutrinos' interaction with background matter,
$U^{\mu}= (U^0, \mathbf{U})= \text{diag} \left( U_{\nu_{e}}^{\mu},U_{\nu_{\mu}}^{\mu},U_{\nu_{\tau}}^{\mu} \right)$, by means of the matrix transformation, $(g_{ab}^{\mu})=\mathcal{U}^{\dagger}U^\mu\mathcal{U}$, where $\mathcal{U}$
is the mixing matrix. The zeroth component, $U^{0}$, is proportional to the
effective number density of background fermions, and the spatial components,
$\mathbf{U}$, are the linear combination of the matter velocity and
polarization. The explicit form of $U^{\mu}$ can be found in Ref.~\cite{DvoStu02}.

The nonperturbative analysis of the complete system~(\ref{eq:Lagrextfields})
which includes the nonzero, nondiagonal interaction with the background matter
and electromagnetic field, is difficult to perform. Nevertheless, it can be
carried out in frames of the perturbation theory. Thus, first, we have to
study the dynamics which results from the diagonal terms of the Lagrangian in Eq.~(\ref{eq:Lagrextfields}). For the sake of the simplification of notations, we
shall omit the index $a$: $\eta \equiv \eta_{a}$, $g^{\mu}\equiv
g_{aa}^{\mu}$, etc. Using the results of Sec.~\ref{sec:CLASS},
we obtain for the new fields an analog of wave equations in Eq.~(\ref{eq:Weyleqetapi}):
\begin{align}\label{eq:etapiextfield}
  \sigma_{\mu}\partial^{\mu}\eta -\mathrm{i}m\sigma_{2}\pi +
  \mathrm{i}g^{\mu}\sigma_{\mu}\eta & =0,
  \notag
  \\
  \sigma_{\mu}^{\ast}\partial^{\mu}\pi -\mathrm{i}m\sigma_{2}\eta -
  \mathrm{i}g^{\mu}\sigma_{\mu}^{\ast}\pi & =0.
\end{align}

Similarly to Sec.~\ref{sec:CLASS}, we introduce the causal propagators $S_{\mathrm{c}}(x-y)=\langle 0|T\{\Xi (x)\Xi^{\dagger}(y)\}|0\rangle $, where $\Xi^{\mathrm{T}}=(\pi ,\eta )$, and the modified propagator $\tilde{S}_\mathrm{c}=-\mathfrak{S}_{2}S_{\mathrm{c}}\mathfrak{S}_{2}\gamma^{0}\gamma
^{5}$. Using Eq.~(\ref{eq:etapiextfield}), we get the equation for the
modified propagator $\tilde{S}_\mathrm{c}$ in the following form:
\begin{equation}\label{eq:tildeSextf}
  \left[
    \Gamma^{\mu}
    \left(
      \mathrm{i}\partial_{\mu}+\mathrm{i}g_{\mu}\Gamma^{5}
    \right) -
    m\Gamma^{5}
  \right]
  \tilde{S}_\mathrm{c} =
  \delta^{4}(x).
\end{equation}
Note that Eq.~(\ref{eq:tildeSextf}) is analogous to one obtained in Ref.~\cite{GeyGitSha00} for the propagator of a massive Dirac particle
interacting with an effective torsion field $S_{\mathrm{eff}}^{\mu}$ if we
make the replacement $g^{\mu}\rightarrow -S_{\mathrm{eff}}^{\mu}$.

Then the path integral representation for $\tilde{S}_{\mathrm{c}}$ can be
found using the results of Ref.~\cite{GeyGitSha00}. Such a representation
has the form
\begin{equation}\label{eq:tildeSpathint}
  \tilde{S}_{\mathrm{c}} =
  \left.
    \exp
    \left(
      \mathrm{i}\Gamma^{n}\frac{\partial_{l}}{\partial \theta^{n}}
    \right)
    Z(\theta^{0},\dots ,\theta^{5})
  \right\vert_{\theta =0},
\end{equation}
where $\theta^{n}$ are Grassmann-odd variables that anticommute with $\Gamma^{n}$, and
\begin{align}\label{eq:propMajextf}
  Z = &
  \int_{0}^{\infty}\mathrm{d}e_{0}
  \int \mathrm{d}\chi_{0}
  \int_{e_{0}} \mathcal{M}(e)\mathrm{D}e
  \notag
  \\
  & \times
  \int_{x_{\mathrm{in}}}^{x_{\mathrm{out}}}\mathrm{D}x
  \int \mathrm{D}\tilde{\pi}
  \int \mathrm{D}\nu
  \int_{\psi (0)+\psi(1)=\theta}\mathcal{D}\psi
  \notag
  \\
  & \times
  \exp
  \left\{
    \mathrm{i}
    \left(
      S_{\mathrm{cl}}+S_{\mathrm{GF}}
    \right) +\psi_{n}(1)\psi^{n}(0)
  \right\}.
\end{align}
Here
\begin{align}\label{eq:ActclMaj}
  S_{\mathrm{cl}} = &
  \int_{0}^{1}
  \bigg[
    -\frac{z^{2}}{2e}-\frac{e}{2}M^{2}+\dot{x}_{\mu}d^{\mu}
    \notag
    \\
    & +
    \mathrm{i}\chi
      \left(
        m\psi^{5}+\frac{2}{3}\psi^{\mu}d_{\mu}
      \right) -
    \mathrm{i}\psi_{n}\dot{\psi}^{n}
  \bigg] \mathrm{d}\tau ,
  \notag
  \\
  M^{2} = & m^{2}+g^{2}+16\partial_{\mu}g^{\mu}\psi^{0}\psi^{1}\psi^{2}\psi^{3},
  \notag
  \\
  z^{\mu} = & \dot{x}^{\mu}+\mathrm{i}\chi \psi^{\mu},
  \quad
  d_{\mu} = 2\mathrm{i}\varepsilon_{\mu \nu \alpha \beta}g^{\nu}\psi^{\alpha}\psi^{\beta},
\end{align}
where $\varepsilon_{\alpha \beta \lambda \sigma}$ is the totally
antisymmetric tensor with $\varepsilon_{0123}=1$, the values $\psi^{n}$ are
Grassmann-odd variables, the term
\begin{equation}\label{Sgf}
  S_{\mathrm{GF}} =
  \int_{0}^{1}
  \left[
    \tilde{\pi} \dot{e}+\nu \dot{\chi}
  \right]
  \mathrm{d}\tau
\end{equation}
is the gauge-fixing action, and the measure $\mathcal{M}(e)$ reads
\begin{equation}\label{eq:measure}
  \mathcal{M}(e) =
  \int \mathcal{D}p\exp
  \left[
    \frac{\mathrm{i}}{2}
    \int_{0}^{1}ep^{2}\mathrm{d}\tau
  \right].
\end{equation}
In Eqs.~(\ref{eq:propMajextf})-(\ref{eq:measure}), $x(\tau )$, $p(\tau )$,
and $\tilde{\pi} (\tau )$ are even trajectories, whereas $\chi (\tau )$ and $\nu (\tau )$ are odd ones and $e$ is the even variable.

One can interpret $S_{\mathrm{cl}}$\ as the parametrization-invariant
pseudoclassical action of a massive neutrino moving in the background
matter. It should be noted that the first example of such action
was the pseudoclassical action of the massive Dirac particle in $3+1$ dimensions
presented by Berezin and Marinov in Ref.~\cite{BM}.

It should be noted that the path integral representation of the propagator in Eqs.~(\ref{eq:tildeSpathint})-(\ref{eq:measure}) could be very useful in calculating
the propagator of a massive Weyl field in background matter since it
accounts for all the loop corrections. Moreover, here we do not restrict
ourselves to the studies of homogeneous matter with characteristics like
number density constant in time, as was assumed in Refs.~\cite{PivStu05,Dvo2010}.

It is interesting to discuss the limiting case $m\rightarrow 0$ of a
massless Weyl particle propagating in background matter. Considering such
a limit in Eq.~(\ref{eq:ActclMaj}), we obtain the following pseudoclassical
action:
\begin{align}\label{eq:ActclMaj0}
  S_{\mathrm{cl}}^{(0)} = &
  \int_{0}^{1}
  \bigg[
    -\frac{1}{2e}(z_{\mu}-ed_{\mu}+eg_{\mu})^{2}
    \notag
    \\
    & +
    z^{\mu}g_{\mu} -
    8e\partial_{\mu}g^{\mu}\psi^{0}\psi^{1}\psi^{2}\psi^{3}
    \notag
    \\
    & -
    \frac{\mathrm{i}}{3}\chi \psi^{\mu}d_{\mu} -
    \mathrm{i}\psi_{n}\dot{\psi}^{n}
  \bigg] \mathrm{d}\tau.
\end{align}

Let us compare Eq.~(\ref{eq:ActclMaj0}) with the pseudoclassical action of a
massless Weyl particle proposed in Ref.~\cite{GitGonTyu94}:
\begin{align}\label{eq:ActclMaj0vac}
  S_{\mathrm{vac}}^{(0)} = &
  \int_{0}^{1}
  \bigg[
    -\frac{1}{2e}
      \left(
        z_{\mu}-\varepsilon_{\mu \nu \alpha \beta}b^{\nu}\psi^{\alpha}\psi^{\beta}+
        \frac{\mathrm{i}}{2}\alpha b_{\mu}
      \right)^{2}
      \notag
      \\
      & -
    \mathrm{i}\psi_{\mu}\dot{\psi}^{\mu}
  \bigg]
  \mathrm{d}\tau,
\end{align}
where $\alpha $ and $b^{\mu}$ are even variables. The quantization of the
action in Eq.~(\ref{eq:ActclMaj0vac}) reproduces the quantum theory of a Weyl
particle. Namely, one gets the description of a right $(\alpha = 1)$ or
left $(\alpha = -1)$ massless neutrino in terms of a bispinor $\Psi_0(x)$,
which satisfies both the Dirac equation with zero mass and the Weyl
condition
\begin{equation}\label{Dm0}
  \mathrm{i}\partial_{\mu}\gamma^{\mu}\Psi_0(x) = 0,
  \quad
  \left(
    \gamma^{5}-\alpha
  \right)
  \Psi_0(x)=0.
\end{equation}
On the contrary, the quantization of the action in Eq.~(\ref{eq:ActclMaj0}) would
recover the description of a massless $1/2$-spin fermion in terms of two-component Weyl spinors $\eta_0$ [see Eqs.~(\ref{Lagreta0}) and~(\ref{eq:WeylLagrm0})]. Note that some alternative pseudoclassical actions for
massless fermions were discussed in Ref.~\cite{Konst}.

Let us choose in Eq.~(\ref{eq:ActclMaj0vac}) the gauge where $\alpha =-1$
(left neutrinos) and $b_{\mu}=2\mathrm{i}eg_{\mu}$. Accounting for the fact that $\psi^{5}=$\textrm{const} if $m=0$, we can cast Eq.~(\ref{eq:ActclMaj0}) in
the form $S_{\mathrm{cl}}^{(0)}=S_{\mathrm{vac}}^{(0)}+S_{\mathrm{int}}^{(0)}$, where
\begin{align}\label{eq:ActclMaj0int}
  S_{\mathrm{int}}^{(0)} = &
  \int_{0}^{1}
  \bigg[
    z^{\mu}g_{\mu}-8e\partial_{\mu}g^{\mu}\psi^{0}\psi^{1}\psi^{2}\psi^{3}
    \notag
    \\
    & -
    \frac{\mathrm{i}}{3}\chi \psi^{\mu}d_{\mu}
  \bigg]
  \mathrm{d}\tau
\end{align}
can be regarded as the pseudoclassical action that describes the interaction
of massless Weyl neutrinos with the background matter.

Finally, let us discuss how to find path integral representations for the
propagators of a massive Weyl field $S_{1,2}$, defined in Sec.~\ref%
{sec:CLASS}. For this purpose we can use Eq.~(\ref{eq:S12SVT}). Let us
decompose the generating function $Z$ in Eq.~(\ref{eq:ActclMaj0int}) in the
variables $\theta$:
\begin{equation}\label{eq:ZGrvar}
  Z =
  \sum_{n=0}^{5}\frac{f_{i_{0}\dots i_{n}}}{n!}
  \theta^{i_{0}} \cdots \theta^{i_{n}},
\end{equation}
where $f_{i_{0}\dots i_{n}}$ are the coefficients, which are antisymmetric
tensors at $n>1$. For $n=0$, $f_{0} = \mathrm{const}$, and $n=1$, the
quantity $f_{n}$ is a vector in a five-dimensional space.

Using Eqs.~(\ref{eq:tildeSpathint}) and~(\ref{eq:ZGrvar}), we represent $\tilde{S}_{\mathrm{c}}$ in the following form:
\begin{align}\label{eq:tildeSSigma}
  \tilde{S}_{\mathrm{c}} = &
  \sum_{n=0}^{5}\frac{\mathrm{i}^{n}}{n!}f_{i_{0}\dots
  i_{n}}\Sigma^{i_{0}\dots i_{n}},
  \notag
  \\
  \Sigma^{i_{0}\dots i_{n}} = & \frac{1}{n!}\Gamma^{\lbrack i_{0}}\dots \Gamma^{i_{n}]}.
\end{align}
On the basis of Eq.~(\ref{eq:tildeSSigma}), we can find the coefficients in
the expansion of $\tilde{S}_{\mathrm{c}}$ [Eq.~(\ref{Scexp})] as follows:
\begin{align}\label{eq:SPVATf}
  S = & f_{0}-\mathrm{i}f_{01235},
  \quad
  P=f_{5}+\frac{\mathrm{i}}{4!} \varepsilon^{\mu \nu \lambda \sigma}f_{\mu \nu \lambda \sigma},
  \notag
  \\
  V_{\mu} = & -f_{\mu 5}-\frac{\mathrm{i}}{3!}
  \varepsilon^{\alpha \beta \lambda \sigma}f_{\alpha \beta \lambda}\eta_{\sigma \mu},
  \notag
  \\
  A_{\mu} = & f_{\mu}+\frac{\mathrm{i}}{3!}
  \varepsilon^{\alpha \beta \lambda \sigma}f_{\alpha \beta \lambda 5}\eta_{\sigma \mu},
  \notag
  \\
  T_{\mu \nu} = & \frac{1}{2}\left( \tilde{f}_{\mu \nu 5}-\mathrm{i}f_{\mu \nu}\right),
  \notag
  \\
  \tilde{f}_{\mu \nu 5} = & -\frac{1}{2}\eta_{\mu \lambda}\eta_{\nu \sigma}
  \varepsilon^{\lambda \sigma \alpha \beta}f_{\alpha \beta 5},
\end{align}
where we have used the representations of the propagators $S_{1,2}$ in Eq.~(\ref{eq:S12SVT}). Here $\eta_{\mu \nu}=\mathrm{diag}(1,-1,-1,-1)$ is the
metric tensor in the Minkowski space. Then, on the basis of Eqs.~(\ref{eq:S12SVT}) and~(\ref{eq:SPVATf}) we can get path integral representations
for $S_{1,2}$.

\section{Massive Dirac neutrinos in external fields\label{sec:DIRAC}}

In Secs.~\ref{sec:CLASS} and~\ref{sec:MAJINTMAT}, we studied both free
massive Weyl field and the interaction of such a field with different backgrounds.
The most probable implementations of such systems are massive Majorana
neutrinos. However, as was mentioned in Sec.~\ref{sec:INTR}, one cannot
exclude the possibility that neutrinos are Dirac particles. That is why we
consider below neutral massive Dirac particles possessing anomalous magnetic
moments, which interact with background matter and with an external
electromagnetic field.

Using the results of Sec.~\ref{sec:FERMEXTF}, we can write down the most
general Lagrangian which describes the interaction of several massive Dirac
particles $\psi_{a}$ with matter and an electromagnetic field. On the basis
of Eq.~(\ref{Lintgen}) we should identify $T^{\mu\nu}_{ab}$ with $\mu_{ab}
F^{\mu\nu}$, where $(\mu_{ab})$ are anomalous magnetic moments and $F_{\mu
\nu} = (\mathbf{E},\mathbf{B})$ is the tensor of the electromagnetic field.
The interaction with background matter can be introduced if we set $V^\mu_{ab} = - A^\mu_{ab} = g^\mu_{ab}/2$ in Eq.~(\ref{Lintgen}), where the
quantities $g_{ab}^{\mu}$ have were already introduced in Eq.~(\ref{eq:Lagrextfields}). Finally, we get this Lagrangian in the following form
(see also Ref.~\cite{Dvo12PAN} and references therein):
\begin{align}\label{eq:LagrDir}
  \mathcal{L} = &
  \bar{\psi}_{a}(\mathrm{i}\gamma^{\mu}\partial_{\mu}-m_{a})\psi_{a} -
  g_{ab}^{\mu}\bar{\psi}_{a}\gamma_{\mu}^{\mathrm{L}}\psi_{b}
  \notag
  \\
  & -\frac{\mu_{ab}}{2}\bar{\psi}_{a}\sigma^{\mu \nu}\psi_{b}F_{\mu \nu},
\end{align}
where $\gamma_{\mu}^{\mathrm{L}}=\gamma^{\mu}(1-\gamma^{5})/2$ .

As we already mentioned in Sec.~\ref{sec:FERMEXTF}, the matrix of magnetic
moments, $(\mu_{ab})$, is Hermitian for the Dirac neutrinos. It means that
nonzero, diagonal elements of this matrix are possible~\cite{Nie82}. The
diagonal element $\mu_{aa}$ corresponds to a magnetic moment of the mass
eigenstate $\psi_{a}$.

The exact analysis of the dynamics of the system described by the Lagrangian
in Eq.~(\ref{eq:LagrDir}) with nonzero, nondiagonal elements of the matrices $(g_{ab}^{\mu})$ and $(\mu_{ab})$ is complicated but can be carried out in
frames of the perturbation theory. That is why, as in Sec.~\ref{sec:MAJINTMAT}, we will discuss only diagonal terms in Eq.~(\ref{eq:LagrDir}) with $a=b$. In this case, we get the wave equation for the propagator of a
single Dirac neutrino $S_{\mathrm{D}}$:
\begin{multline}
  \bigg[
    \mathrm{i}\gamma^{\mu}\partial_{\mu}-m-\frac{1}{2}\gamma^{\mu}(1-\gamma^{5})g_{\mu}
    \\
    -
    \frac{\mu}{2}\sigma_{\mu \nu}F^{\mu \nu}
  \bigg]
  S_{\mathrm{D}}(x) =
  -\delta^{4}(x).  \label{eq:Dirnuwe}
\end{multline}
where $m\equiv m_{a}$, $g^{\mu}\equiv g_{aa}^{\mu}$, and $\mu \equiv
\mu_{aa}$. Then, using Eq.~(\ref{eq:Dirnuwe}), we obtain the following
equation for the modified propagator, $\tilde{S}_{\mathrm{D}}=S_{\mathrm{D}}\Gamma^{5}$:
\begin{multline}\label{eq:Dirnuwemod}
  \bigg[
    \Gamma^{\mu}
    \left(
      \mathrm{i}\partial_{\mu}-\frac{1}{2}g_{\mu}+\frac{\mathrm{i}}{2}\Gamma^{5}g_{\mu}
    \right)
    -m\Gamma^{5}
    \\
    -\frac{\mathrm{i}\mu}{2}\Gamma^{5}\Gamma^{\mu}\Gamma^{\nu}F_{\mu \nu}
  \bigg]
  \tilde{S}_{\mathrm{D}}(x)=\delta^{4}(x),  
\end{multline}
where the matrices $\Gamma^{\mu}$ and $\Gamma^{5}$ were defined in Sec.~\ref{sec:CLASS}.

One can see that the vector part of the interaction with matter is
equivalent to the interaction with an effective electromagnetic field: $(qA^{\mu})_{\mathrm{eff}}=g^{\mu}/2$, where $q$ is an effective electric
charge. The axial vector part of the interaction with matter is equivalent
to the interaction with an effective torsion field: $S_{\mathrm{eff}}^{\mu}=-g^{\mu}/2$. Contributions due to such interactions to the propagator of
the Dirac massive particle were studied in Ref.~\cite{GeyGitSha00}. The
analysis of the propagator of a fermion with an anomalous magnetic moment
was made in Ref.~\cite{Git97}.

Using the results of the latter works, we can obtain the path integral
representation for the propagator $\tilde{S}_{\mathrm{D}}$. The general
structure of such a representation is similar to one given by Eqs.~(\ref{eq:tildeSpathint}), (\ref{eq:propMajextf}), (\ref{Sgf}), and~(\ref{eq:measure}) with the substitution of the pseudoclassical action in Eq.~(\ref{eq:ActclMaj}) by the following one:
\begin{align}\label{eq:ActDirac}
  S_{\mathrm{cl}} = &
  \int_{0}^{1}
  \bigg[
    -\frac{z^{2}}{2e}-\frac{e}{2}M_{\mathrm{D}}^{2}-\frac{\dot{x}_{\mu}}{2}
    \left(
      g^{\mu}-d^{\mu}+8\mathrm{i}\mu \psi^{5}F^{\mu \nu}\psi_{\nu}
    \right)
    \notag
    \\
    & +
    \mathrm{i}\frac{e}{2}G_{\mu \nu}\psi^{\mu}\psi^{\nu}
    \notag
    \\
    & +
    \mathrm{i}\chi
    \left(
      m_{\mathrm{D}}^{\ast}\psi^{5}+\frac{1}{3}\psi^{\mu}d_{\mu}
    \right) -\mathrm{i}\psi_{n}\dot{\psi}^{n}
  \bigg]
  \mathrm{d}\tau,
\end{align}
where
\begin{align*}
  G_{\mu \nu} = & \partial_{\mu}g_{\nu}-\partial_{\nu}g_{\mu},
  \\
  M_{\mathrm{D}}^{2} = & m_{\mathrm{D}}^{2}+g^{2}/4+8\partial_{\mu}g^{\mu}\psi^{0}\psi^{1}\psi^{2}\psi^{3},
\end{align*}
and $m_{\mathrm{D}}=m-2\mathrm{i}\mu F_{\alpha \beta}\psi^{\alpha}\psi^{\beta}$.

Again, we can mention that Eqs.~(\ref{eq:tildeSpathint}), (\ref{eq:propMajextf}), and~(\ref{eq:ActDirac}) are the most complete expression
for a propagator of a massive neutral Dirac particle with an anomalous magnetic
moment, since they exactly account for the influence of background matter
and an external electromagnetic field. Besides the loop corrections, these
expressions also contain the inhomogeneous matter contributions, since we do
not suppose that $\partial_{\mu}g_{\nu}=0$. Thus, Eqs.~(\ref{eq:propMajextf}), (\ref{eq:tildeSpathint}), and~(\ref{eq:ActDirac})
generalize the results of Refs.~\cite{PivStu05,Dvo2010}, where the neutrino
propagator in homogeneous matter was obtained on tree level without loop
corrections.

\section{Conclusion\label{sec:CONCL}}

In the present work, we have studied massive $1/2$-spin particles in various
external backgrounds, keeping in mind applications to neutrino physics. We
have been mainly interested in massive Majorana (or Weyl) fields. However,
massive neutral Dirac particles have also been considered. We have
formulated classical Lagrangian theory of the massive Weyl field in terms of
Grassmann-odd two-component spinors. Then, we have constructed the
Hamiltonian formulation of such a theory, which turns out to be a theory
with second-class constraints. Using this formulation, we have canonically
quantized the massive free Weyl field. We have derived propagators of the
Weyl field and related them to the propagator of a massive Dirac particle.

Then we have studied the massive Weyl particles propagating in a background
mater. We have found the path integral representation for the propagator of
the massive Weyl field in background matter, as well as we have obtained
the corresponding pseudoclassical action for massive Weyl particles. The
massless limit of such an action was compared with the results of other works.
Finally, we have studied the path integral representations for a neutral
massive Dirac particle with an anomalous magnetic moment moving in
background matter under the influence of an external electromagnetic field.
From this representation, we have derived the pseudoclassical action of the
corresponding neutral massive Dirac neutrino with an anomalous magnetic
moment.

The results of the present work can be applied for the study of the
propagation of massive mixed (Majorana or Dirac) neutrinos in dense
matter and strong electromagnetic fields. As demonstrated in Ref.~\cite{GitZlat}, path integral representations of particle propagators allow
one to effectively calculate the propagators. We hope that the
representations derived in the present work will be helpful for describing
the neutrino motion in various astrophysical and cosmological media (see,
e.g., Ref.~\cite{Raf96}).

In the case of a Dirac neutrino moving in a singular external background one
should carefully define boundary conditions for corresponding solutions,
as this is equivalent to choosing a self-adjoint Hamiltonian in such a
background. To this end, one can use a general theory of self-adjoint
extensions of symmetric operators and some analogy with the motion of charged
particles in an Aharonov-Bohm field (see, e.g., Ref.~\cite{book2}).

It should be noted that neutrinos may have magnetic moments (for a review,
see Refs.~\cite{Dvo2010,BroGiuStu12}). The best experimental constraint on
the Dirac magnetic moment of an electron neutrino is $\mu_{\nu_{e}} < 2.9\times 10^{-11}\mu_{\mathrm{B}}$~\cite{Bed12}, where $\mu_{\mathrm{B}}=e/2m_{e}$ is the Bohr magneton. Astrophysical constraints on the
Dirac magnetic moments are stronger: $\mu_{\nu}<1.1\times 10^{-12}\mu_{\mathrm{B}}$~\cite{KuzMikOkr09}. Note that in order to satisfy $m_{\nu}\lesssim 1\,\text{eV}$, there should be a more natural scale for the Dirac
magnetic moments of a neutrino: $\mu_{\nu}\lesssim 10^{-14}\mu_{\mathrm{B}}$~\cite{Bell05}. If neutrinos have magnetic moments, an inhomogeneous
magnetic field, acting on such particles, can change their kinetic energy,
i.e. produce work. This fact implies the possibility of the neutrino
creation form the vacuum by strong inhomogeneous magnetic fields (see, e.g.,
Ref.~\cite{FraGitShv91}). In the case of Dirac neutrinos, this effects was
studied in Ref.~\cite{GavGit12}. In this connection, it is interesting to
generalize the technique of the present work to quantize Majorana neutrinos
that have anomalous transition magnetic moments, and then to study the
creation of such neutrinos by strong inhomogeneous magnetic fields. In
addition, by applying similar methods, one can study the creation of
Majorana neutrinos from vacuum by the inhomogeneous background matter.

%\section*{Acknowledgments}

\begin{acknowledgments}
We are grateful to I.~V.~Tyutin for helpful discussions. M.D. is thankful to
FAPESP for supporting his postdoctoral position. D.M.G. thanks FAPESP and CNPq
for permanent support and the Russian Ministry of Education and Science for
support under the Project 14.B37.21.0911.
\end{acknowledgments}

\end{document}